\documentclass[jmp]{revtex4-1}
\usepackage{amsmath, amssymb}
\usepackage{graphicx}
\usepackage{dcolumn}
\usepackage{bm}
\usepackage{blindtext}
\usepackage[utf8]{inputenc}
\usepackage[T1]{fontenc}
\usepackage{mathptmx}
\usepackage{etoolbox, color}

\makeatletter
\def\@email#1#2{%
 \endgroup
 \patchcmd{\titleblock@produce}
  {\frontmatter@RRAPformat}
  {\frontmatter@RRAPformat{\produce@RRAP{*#1\href{mailto:#2}{#2}}}\frontmatter@RRAPformat}
  {}{}
}%
\date{}
\makeatother
\begin{document}
\preprint{AIP/123-QED}

\title{A Quantum Approach to the Continuum Heisenberg Spin-Chain Model: Position-Dependent Mass Formalism and Pre-canonical Quantization}
 \author{V. Chithiika Ruby}
 \email{Corresponding Author: chithiikaruby.v@trp.srmtrichy.edu.in}
  \affiliation{Center for Research, Easwari Engineering College, Chennai-600 089, Tamil Nadu, India.}
  \affiliation{Center for Nonlinear and Complex Networks, SRM TRP Engineering College, Tiruchirappalli-621 105, Tamil Nadu, India.}
 \author{M. Lakshmanan}%
 \email{lakshman.cnld@gmail.com}
 \affiliation{Department of Nonlinear Dynamics, School of Physics, 
  Bharathidasan University, Tiruchirappalli - 620 024, Tamil Nadu, India.}

\date{\today}

\begin{abstract}
Painlev\'{e}'s singularity structure analysis, combined with stereographic mapping, has previously been applied to a one-dimensional Heisenberg spin-chain continuum model which identified a Hamiltonian density for the static version of the Landau-Lifshitz equation. In this work, we explore the equivalence of the Hamiltonian density to the nonlinear sigma model. It reveals its non-standard form and can be interpreted as a position-dependent mass Hamiltonian density. We then proceed with the quantization of this Hamiltonian density using the pre-canonical quantization procedure.  The resulting Schr\"{o}dinger-like equation was found to take the form of a confluent Heun equation. By employing the functional Bethe-Ansatz method, we explicitly obtain the ground state and first excited state of the system. This analysis provides a comprehensive quantum description of the system, capturing the probabilistic structure of the field and information about the possible energy states of the spin system.

\end{abstract}

\maketitle

\section{Introduction}
Spin chains, as one-dimensional arrays of interacting spins, have emerged as primary systems for exploring fundamental phenomena in statistical physics \cite{Gia}, condensed matter physics \cite{auerbach2012interacting}, and classical as well as quantum field theory \cite{francesco2012conformal, fradkin2013field}.  The continuum formulation of spin chains typically involves replacing the discrete spin variables with continuous field variables which could be scalar $\mathbf{\phi}(x)$, spinor $\mathbf{\psi}(x)$, or vector fields $\mathbf{A}(x)$.  This approach captures the low-energy, long-wavelength behavior of spin chains at macroscopic level \cite{takhtajan1982picture}. One such model is the nonlinear sigma model (NLSM), where spins are represented by vector fields constrained to lie on a unit sphere.  Nonlinear models, including the NLSM, remain indispensable tools for understanding the phenomena such as low-energy excitations in quantum spin chains \cite{affleck1988universal}, macroscopic magnetic properties like magnetization and magnetic susceptibility \cite{haldane1983nonlinear, landau1935theory}, and topologically nontrivial configurations such as skyrmions \cite{skyrme1961non, skyrme1994non}. In addition, the application of functional renormalization group methods on the continuum models reveals their connection to critical phenomena \cite{efremov2021nonlinear}. It is also fundamental to the study of solitons in spin-chain systems with weak easy-axis anisotropy  \cite{haldane1983nonlinear, daniel1992singularity} and to the study of interacting Fermi gas as a quantum liquid model \cite{luther1974backward, haldane1994luttinger}. 
During 1992, M. Daniel et al studied the integrable/non-integrable property of the Heisenberg ferromagnetic spin chain Hamiltonian with uniaxial  in external magnetic field, $\bf{B(t)}$, \cite{daniel1992singularity}
\begin{equation}
{\cal H} = - J \sum_i {\bf {S}}_i.{\bf {S}}_{i+1} + A \sum_i (S^z_i)^2 - \mu {\bf {B}}. \sum_i {\bf {S}}_i, \label{hamiltonian1}
\end{equation}
in the continuum limit, where ${\bf {S}}_i = S^{x}_i \hat{e}_x + S^{y}_i \hat{e}_y +S^{z}_i \hat{e}_z, \; i = 1, 2, 3, ....N,$ with $|{\bf S}|^2 = 1$ and the parameter $J$ denotes the pair-interaction $(J > 0)$ and $\mu = g \mu_B$ is the gyromagnetic ratio. The anisotropy parameter $A$ is positive along the easy plane, whereas $A < 0$ along easy axis. The one-dimensional classical Heisenberg ferromagnetic spin chain is considered to be an interesting nonlinear dynamical system as it exhibits both coherent and chaotic structures depending on the nature of the applied magnetic field. To analyze the integrability of the system, they used a transformation on the  Landau-Lifshitz spin field equation corresponding to the system (\ref{hamiltonian1}), which involves a stereographic projection of the spin vector onto a complex plane introduced in Refs. \cite{daniel1983perturbation, lakshmanan1984landau}.
The Landau-Lifshitz equation derived for the system (\ref{hamiltonian1}) by utilizing the Poisson bracket relations for the spin components takes the following form in the continuum limit:
\begin{equation}
\frac{\partial \mathbf{S}}{\partial t} = \mathbf{S} \times \left[\mathbf{S}_{zz} - 2 A (\mathbf{S}.\mathbf{n})\mathbf{n} + \mu \mathbf{B}\right],
\label{LL-equation1}
\end{equation}

where $ \mathbf{S} $ is the spin vector and  $ \mathbf{B} $ is the external magnetic field.

\subsection{Stereographic projection}

In quantum mechanics, the Bloch sphere representation is a geometric visualization used to represent the state of a two-level quantum system, such as a spin-$\frac{1}{2}$ particle. This maps the quantum state of the spin onto a point on the surface of a sphere, providing a framework to understand spin dynamics and superposition. Hence a spin vector is represented as a radial vector of the sphere with $\mathbf{S}^2 = 1$. In the study \cite{daniel1983perturbation}, it is highlighted that stereographically projecting the spin vector onto a complex plane simplifies the analysis of the spin field equation in the presence of anisotropy.  Moreover, the quantum state of the spin is considered to be mapped onto a sphere of radius $R$, where $\mathbf{S}^2 = R = 1$ (without loss of generality), and subsequently projected onto an extended complex plane. It is expressed as, 
\begin{equation}
\omega(z, t) = \frac{S_1 + i S_2}{1 + S_3}, \label{stereo-mapping}
\end{equation} 
where ${\bf S}(z, t) = (S_1, S_2, S_3)$, whose inversion leads to 
\begin{equation}
S_1 + i S_2 = \frac{2\omega}{1 + \omega \omega^*}, \qquad S_3 = \frac{1  - \omega\omega^*}{1  + \omega\omega^*}. \label{stereo-mapping-inversion}
\end{equation} 
When stereographically projecting the sphere onto the complex plane, the north and south poles map to the points at infinity and origin, respectively. Any other point on the sphere corresponds to a complex number in the complex plane, where the state can be written in terms of a single complex parameter.

The integrability of the system (\ref{hamiltonian1}), in the continuum limit, has been analyzed through the spin field equation after performing a stereographic projection of the spin vector onto the complex plane \cite{daniel1992singularity}. The equation is transformed to the following form with $\omega$ as, 
\begin{eqnarray}
i(1 + \omega \omega^*)\omega_t + (1 + \omega \omega^*) \omega_{zz}  - 2 \omega^* (\omega_z)^2 + 2 A \omega (1 - \omega \omega^*) + \mu (1 + \omega \omega^*)  \left[\frac{B^x}{2} (1 - \omega^2) + \frac{B^y}{2} (1 + \omega^2) - B^z\omega\right]= 0, 
\label{LL-equation2}
\end{eqnarray}
where the external magnetic field ${\bf B} = (B^x, B^y, B^z)$.

By employing the Painlev\'{e} singularity structure analysis, it was shown that the system is integrable in the absence of $A$ or magnetic field ${\bf B}$. While analyzing the static version of the field equation Eq.(\ref{LL-equation2}), say $\omega(z, t) = P(z) + i Q(z)$, the associated Hamiltonian density  is obtained as 
\begin{equation}
\mathcal{H}(z) =  \frac{1}{2}(1 + P^2+ Q^2)^2\;(\Pi^2_P + \Pi^2_Q) + V(P, Q), 
\label{hamil2}
\end{equation}
where the conjugate spatial momenta are defined as,  
\begin{equation}
\Pi_P =  \frac{P_z}{(1 + P^2 + Q^2)^2}, \quad \Pi_Q = \frac{Q_z}{(1+P^2+Q^2)^2}, 
\label{momenta}
\end{equation}
where the subscript $z$ denotes differentiation with respect to $z$. And the potential $V(P, Q)$ is of the form, 
\begin{eqnarray}
V(P, Q) = -\frac{A}{4}\frac{(1 - P^2 - Q^2)^2}{(1 +  P^2 + Q^2)^2} +\frac{\mu B}{2}\frac{P}{(1 +  P^2 + Q^2)}, 
\end{eqnarray}
 where $B = B^x$ is the strength of the constant magnetic field and $B^y = B^z =0$ \cite{daniel1992singularity}.

\subsection{Nonlinear Sigma Model} 
Before quantizing the Hamiltonian density (\ref{hamil2}), we first examine its realization as a nonlinear sigma model. 
The Lagrangian density for a set of coupled scalar fields, $\phi_{i}(x^{\mu})$, where $i = 1, 2, \dots, D$, in a $d$-dimensional flat space-time with indices, $\mu = 0, 1, 2, \dots, d-1$, is given by 
\begin{eqnarray}
\hspace{-5cm}\mathcal{L} &=& \frac{1}{2} G_{ij}(\phi) \partial^{\mu} \phi^i \partial_\nu \phi^j, \\
\hspace{-4cm}\mbox{or} \hspace{4cm}\ & & \nonumber \\
\hspace{-5cm}\mathcal{L} &=& \frac{1}{2} g^{\mu\nu}  G_{ij}(\phi) \partial_\nu \phi^i \partial_\nu \phi^j,
\label{nlsm_L}
\end{eqnarray}
where $\partial^{\mu}  =  g^{\mu\nu} \partial_{\nu}$ and $g^{\mu\nu}$  is the space-time metric and $G_{ij}(\phi)$ is the target space metric and it should be positive definite \cite{ketov2013quantum}. This is referred as the nonlinear sigma model, a type of field theory that characterizes the dynamics of a set of scalar fields constrained to a nonlinear manifold, often a symmetric space such as a sphere or a group manifold.

If the target space is a sphere of radius $ R $, and the field $ \phi_i $ is generally constrained to lie on the sphere, i.e., $ \phi_k \phi^k = 1 $, where the index $ k $ is summed over all possible values of $k$ and the target-space metric $G_{ij}(\phi) \approx R^2 \;\delta_{ij}$.

We consider the Lagrangian density for the scalar fields $\phi_{i}(x, y, z) , \; \; i = 1, 2, 3$, defined on a Euclidean space:
\begin{eqnarray}
\mathcal{L}(\phi, \partial_\mu \phi) = \frac{1}{2} \sum_{i=1}^{3} (\partial_\mu \phi_{i})(\partial^\mu \phi_{i}) - V(\phi),
\end{eqnarray}
where $V(\phi)$ is the potential term. We suitably consider a transformation, 
\begin{eqnarray}
\phi(x) &\rightarrow& \phi'(x'),  \label{conformal}\\
x &\rightarrow& x', \label{conforma2}
\end{eqnarray}
which is achieved by $\phi'(x') = \mathcal{F}(\phi(x))$. Under this transformation, the action,
\begin{equation}
S = \int d^{d}x \, {\cal L}(\mathbf{\phi}(x), \partial_{\mu}\mathbf{\phi}(x))
\label{action}
\end{equation} 
is invariant. Here the field \(\phi(x)\), considered as a mapping from space-time to some target space \({\cal M}\) \((\phi: \mathbb{R}^d \rightarrow {\cal M})\), undergoes both functional (\ref{conformal}) and coordinate changes (\ref{conforma2}). Such a transformation is known as a passive transformation. On the other hand, if only the coordinates change (\ref{conforma2}), the transformation is known as an active transformation \cite{francesco2012conformal}.

We consider three independent scalar fields, $S_1(z)$, $S_2(z)$, and $S_3(z)$, that lie on a sphere $(S_1^2 + S_2^2+ S_3^2 = 1)$ and the corresponding Lagrangian density is give in the form: 

\begin{eqnarray}
\mathcal{L} &=& \frac{1}{2}\left[(\partial_z S_1)^2 + (\partial_z S_2)^2 + (\partial_z  S_3)^2\right] - V(S_1, S_2, S_3).\label{lagrangian-Spin}
\end{eqnarray}

The spin field vectors defined on a sphere can be mapped onto the complex plane, where their components are represented as projections onto the real and imaginary axes. This mapping is achieved through a stereographic projection of the sphere onto the complex plane. The unitary transformation implementing this stereographic projection is expressed as:

\begin{eqnarray}
U = 
\begin{pmatrix}
S_3 & S_1 - i S_2\\
S_1 + i S_2 & -S_3
\end{pmatrix},
\end{eqnarray}
where $U^\dagger U = 1$.

In this formalism, the spin dynamics is described using a complex-valued function, with its real and imaginary parts corresponding to orthogonal components of the spin vector. Based on the complex-plane representation of the spin field vectors 
(\ref{stereo-mapping-inversion}), the unitary transformation is expressed in terms of the stereographic mapping as

\begin{eqnarray}
U = \frac{1}{1 + |\omega|^2}
\begin{pmatrix}
1 - |\omega|^2 & 2 \omega^{*} \\
2 \omega & -1 + |\omega|^2
\end{pmatrix},
\end{eqnarray}
where $\omega$ is the stereographic projection coordinate, and $|\omega|^2$ is the squared modulus of \(\omega\).

The unitary transformation thus changes the Lagrangian density \(\mathcal{L}\) (without potential term) to be
\begin{eqnarray}
\mathcal{L}_{kinetic} &=& \frac{1}{2} Tr[\left(\partial_\mu U \right)^{\dagger}\left(\partial_\mu U\right)], 
\end{eqnarray}
and so we get
\begin{eqnarray}
\mathcal{L}_{kinetic} &=& \frac{2}{(1 + |\omega|^2)^2} \frac{\partial \omega}{\partial z} \frac{\partial \omega^*}{\partial z}, 
\end{eqnarray}

As in the work \cite{daniel1983perturbation} we assume  $\omega = P + i Q$, and so we could get the kinetic energy part as
\begin{eqnarray}
\mathcal{L}_{kinetic} &=& \frac{2}{(1 + P^2 + Q^2)^2} (P^2_z + Q^2_z), 
\label{Lagrangian-nlsm}
\end{eqnarray}
where suffix $z$ denotes the differentiation with respect to $z$. 

Thus, the Hamiltonian presented in equation (\ref{hamil2}) is demonstrated to realize the nonlinear sigma model in the classical regime. It forms the foundation for future studies in both classical and quantum domains. The classical analysis shows that the one-dimensional continuum Heisenberg anisotropic ferromagnetic spin chain in a transverse magnetic field is non-integrable and exhibits chaotic spatial patterns. The system becomes integrable when either the anisotropic interaction or the magnetic field is removed. Therefore, the transverse magnetic field plays a key role in disrupting  the integrability of the system, leading to chaotic behavior and complex dynamics \cite{daniel1992singularity, daniel1983perturbation}. 

In the following sections, we will focus on the quantum mechanical perspective and examine the possibility of exact solutions.

\section{Quantization}
The standard approach in the literature for quantizing the Hamiltonian of a field involves representing the field operator as a superposition of annihilation and creation operators of harmonic oscillator, each corresponding to a mode parametrized by the wave vector \cite{auerbach2012interacting}. These operators govern the creation and annihilation of particles associated with specific modes defined by the wave number.  In the quantum Luttinger liquid framework, it has been demonstrated that the Hamiltonian can be diagonalized using Sturm-Liouville theory. In this formalism, the fields and their conjugate momenta are expressed as expansions in terms of annihilation and creation operators. The expansion coefficients are weighted by eigenfunctions of the Sturm-Liouville operator, which encapsulates the spin interaction Hamiltonian in the continuum limit, where the Luttinger parameter varies with position \cite{moosavi2024perfect}. Alternatively, the \(O(2)\)-nonlinear sigma model has been quantized through the functional Schr\"{o}dinger equation to determine the ground and first excited state wave functional.  In this framework, the global variance is analyzed, which captures the overall evolution and statistical behavior of the system at macroscopic scales \cite{muslih2002quantization}. On the other hand, the local variance of the field in the quantum regime is examined within the context of Clifford algebra, offering a detailed understanding of spatial fluctuations and point-wise interactions at the level of individual spins, known as pre-canonical quantization \cite{kanatchikov1998toward, kanatchikov2001precanonical}.

In this study, we employ the pre-canonical quantization procedure to investigate the quantum dynamics of the Hamiltonian density (\ref{hamil2}).  The pre-canonical quantization method directly works with the field components and their spacetime derivatives, ensuring the preservation of covariance under spacetime transformations. This approach is based on the De Donder-Weyl formulation of classical field theory which  introduces the concept of polymomenta, conjugate momenta associated with the fields \cite{weyl1934observations, rund1966hamilton}. It means that it extends Hamiltonian mechanics into field theory through a multi-temporal  framework, treating space and time symmetrically, in a similar way as time is treated in classical mechanics \cite{kanatchikov1998toward, kanatchikov2001precanonical}. These polymomenta are defined as:
\begin{equation}
\pi^{\mu}_a = \frac{\partial \mathcal{L}}{\partial_{\mu} \phi^a}, \label{polymomenta}
\end{equation}
where $\mathcal{L}$ is the Lagrangian density and $\phi^a(x^{\mu})$ represents the components of the scalar field with $x^{\mu} := (x^t, x^i)$. 

To describe the quantum evolution of these fields, field components and polymomenta are treated as operators as, 
\begin{equation}
\hat{\pi}^{\mu}_a = -i \hbar \kappa \gamma^{\mu} \frac{\partial}{\partial \phi^a},\quad \phi^a = \hat{\phi}^a,  \label{polymomenta_operator}
\end{equation}
where $\kappa$ is a constant and it is related to the inverse of the volume in space dimension and the Dirac matrices, $\gamma^{\mu}$, ensure that the algebra associated with the operators preserves spacetime symmetries \cite{kanatchikov1998toward, kanatchikov2001precanonical}. They satisfy the following commutation relation, 
\begin{equation}
[\hat{\phi}^a, \hat{\pi}^\mu_b] =i \hbar \kappa \gamma^\mu \delta^a_b, \label{commutation_relation1}
\end{equation}
where $\delta^a_b$ is Kronecker delta function which takes the value $1$, if $a = b$ otherwise $0$. These operators act on Clifford-valued wave functions, $\Psi(\phi^a, x_{\mu})$, as 
\begin{equation}
i \hbar \kappa \gamma^{\mu} \partial_{\mu} \Psi = \hat{\mathcal{H}} \Psi.  \label{pre-canonical}
\end{equation}
In the pre-canonical Schr\"{o}dinger-like equation (\ref{pre-canonical}), the term  $\gamma^{\mu}\; \partial_{\mu}$ ensures covariance of space-time and $\hat{\mathcal{H}}$ is the Hamiltonian density operator. 

In this study, the Hamiltonian density describing a static field is expressed as  
\begin{equation}
\mathcal{H}(z) \equiv \mathcal{H}(P, Q, \Pi_P, \Pi_Q) = \frac{1}{2m}(\Pi_P^2 + \Pi_Q^2) - \frac{A}{4} \frac{(1 - P^2 - Q^2)^2}{(1 + P^2 + Q^2)^2} + \frac{\mu B}{2} \frac{P}{(1 + P^2 + Q^2)}, \label{H_PQ}
\end{equation}  
where $P(z)$ and $Q(z)$ are two independent field components that vary with the $z$-coordinate. In this context, the conjugate momenta correspond to the spatial momenta only. The formulation is restricted to purely spatial coordinates (time-independent setup).  

We  define  the field operators, $\hat{P}$ and $\hat{Q}$, along with their corresponding conjugate momenta as,  
\begin{equation}
\hat{\Pi}_P = -i\hbar \frac{\partial}{\partial P}, \quad \qquad \hat{\Pi}_Q = -i\hbar \frac{\partial}{\partial Q}, \label{momenta-operators}
\end{equation}  
where the role of Dirac matrices is absent due to the restriction of dynamics to a single dimension, that is $z$-direction.  

The operators $\hat{P}$, $\hat{Q}$, and their conjugate momenta satisfy the standard commutation relations:  
\begin{equation}
[\hat{P}, \hat{\Pi}_P] = i\hbar, \quad \qquad  [\hat{Q}, \hat{\Pi}_Q] = i\hbar. \label{commutator}
\end{equation}  

For simplicity, the characteristic length scale in Eq.~(\ref{pre-canonical}) is normalized to $\kappa = 1$, as the dynamics is reduced to a single spatial variable ($z$).  

It is important to emphasize that the Hamiltonian density under consideration is non-standard, with the terms involving the mass function $m(P, Q)$ and the momentum terms are intricately interwoven. This structure reflects the underlying complexity of the system and its position-dependent nature, as highlighted in the literature on position-dependent mass Schr\"{o}dinger equations \cite{bastard1990wave, gonul2002exact, mathews1975quantum, higgs1979dynamical}. To continue, we express the Hamiltonian (\ref{H_PQ}) in operator form using (\ref{momenta-operators}) and substitute it into the pre-canonical Schr\"{o}dinger-like equation (\ref{pre-canonical}). Additionally, we adopt the symmetric ordering for the Hamiltonian operators. The resulting equation takes the following form:
\begin{eqnarray}
i \hbar \frac{\partial \Psi}{\partial z} =  \left[\left( \frac{1}{4 m(P, Q)} \left( \hat{\Pi}_P^2 + \hat{\Pi}_Q^2 \right) + \left( \hat{\Pi}_P^2 + \hat{\Pi}_Q^2 \right) \frac{1}{4 m(P, Q)}\right) - \frac{A}{4} \frac{(1 - P^2 - Q^2)^2}{(1 + P^2 +Q^2)^2} +\frac{\mu B}{2}\frac{P}{(1 + P^2 + Q^2)} \right]\Psi, \label{pre-canonical-se}
\end{eqnarray}
where the wave function $\Psi(P, Q, z)$ is locally defined, and the momentum operators $\hat{\Pi}_P$ and $\hat{\Pi}_Q$ correspond to the conjugate momenta of the fields $P(z)$ and $Q(z)$, respectively. The mass function ${\displaystyle m(P, Q) = \frac{1}{(1 + P^2 + Q^2)^2}}$ is position-dependent and its explicit form plays a crucial role in the dynamics of the system. 

It should be emphasized that the non-commutative nature of the canonically conjugate variables in the quantum system admits different orderings of the Hamiltonian (\ref{H_PQ}). In this case, the momentum and mass operators are ordered in a Hermitian way \cite{carinena2004non}. It is also shown that the non-Hermitian ordered form of the Hamiltonian (\ref{H_PQ}) admits real energy eigenvalues due to its connection to the Hermitian part, which is known as the quasi-Hermitian nature \cite{chithiika2015removal}. Moreover, it is demonstrated that the solvability of the quantum system is determined by the factors of the ordering \cite{ruby2024lienard}.

Equation (\ref{pre-canonical-se}) can be solved using the variable separation method and so we assume, 
\begin{equation}
\Psi(P, Q, z) = e^{-\frac{i}{\hbar} E z} \psi(P, Q), \label{sol_pq-z}
\end{equation}
which reduces the equation (\ref{pre-canonical-se}) to the form, 
\begin{equation}
\left[ \frac{1}{4 m(P, Q)} \left( \hat{\Pi}_P^2 + \hat{\Pi}_Q^2 \right) + \left( \hat{\Pi}_P^2 + \hat{\Pi}_Q^2 \right) \frac{1}{ 4\;m(P, Q)} - \frac{A}{4} \frac{(1 - P^2 - Q^2)^2}{(1 + P^2 + Q^2)^2} + \frac{\mu B}{2} \frac{P}{(1 + P^2 + Q^2)^2} \right] \psi(P, Q) = E \psi(P, Q) , \label{pre-canonical-se-pq}
\end{equation}
where  ${\displaystyle m(P, Q) = \frac{1}{(1 + P^2+ Q^2)^2}}$ is the mass function.

Under this transformation,  $P = r \cos \phi$ and $Q = r \sin\phi$, the equation (\ref{pre-canonical-se-pq}) becomes, 
\begin{eqnarray}
\frac{-\hbar^2}{ 4}\left[\frac{1}{m(r)}\nabla^2 + \nabla^2 \frac{1}{m(r)}\right] \psi(r, \phi)+\left[ -\frac{A}{4}\frac{(1- r^2)^2}{(1 + r^2)^2} + \frac{\mu\;B r \cos{\phi}}{2\;(1 + r^2)^2}\right]\psi(r, \phi) = E\psi(r, \phi), \label{se_polar}
\end{eqnarray}
where $ {\displaystyle \nabla^2 =\frac{\partial^2}{\partial r^2} + \frac{1}{r}\;\frac{\partial}{\partial r} + \frac{1}{r^2 } \frac{\partial^2}{\partial \phi^2}}$. 
To analyze the solvability of equation (\ref{se_polar}), we consider $B = 0$ (to start with) and it becomes 
\begin{eqnarray}
\frac{\partial^2 \psi}{\partial r^2} + \left(\frac{4 r}{1+ r^2} + \frac{1}{r}\right) \frac{\partial \psi}{\partial r} + \frac{1}{r^2} \frac{\partial^2 \psi}{\partial \phi^2}+
\left[\frac{8}{1 + r^2}+\frac{\left(\frac{2 E}{\hbar^2}+\frac{A}{2\hbar^2}\right) - 4}{(1+r^2)^2}-\frac{2 A}{\hbar^2}\frac{1}{(1 + r^2)^3}+\frac{2A}{\hbar^2 (1 + r^2)^4}\right]\psi = 0. 
\label{se_r_phi}
\end{eqnarray}
Let us consider the separation of variables $\psi(r, \phi) = \chi(r)\Phi(\phi)$, so that the equation (\ref{se_r_phi}) is separated into angular and radial parts as follows 
\begin{equation}
\frac{\partial^2 \Phi}{\partial \phi^2} = -\lambda^2 \Phi \label{anugular_part}
\end{equation}
and radial part:
\begin{eqnarray}
\frac{d^2 \chi}{d r^2} + \left(\frac{4 r}{1+ r^2} + \frac{1}{r}\right) \frac{d \chi}{d r} +
\left[-\frac{\lambda^2}{r^2}+\frac{8}{1 + r^2}+\frac{\left(\frac{2 E}{\hbar^2}+\frac{A}{2\hbar^2}\right)-4}{(1+r^2)^2}-\frac{2 A}{\hbar^2}\frac{1}{(1 + r^2)^3}+\frac{2A}{\hbar^2 (1 + r^2)^4}\right]\chi = 0.
\label{se_r_phi1}
\end{eqnarray}
On using the transformations,
\begin{equation}
\chi(r) = \frac{r^{\lambda}}{(1 + r^2)^l}\;\exp{\left[\frac{a}{1 + r^2}\right]}\;S(r), \qquad \zeta = \frac{1}{1 + r^2}
\label{transformation}
\end{equation}
with ${\displaystyle a = \frac{\sqrt{A}}{\hbar\sqrt{2}}}$, we can reduce the equation (\ref{se_r_phi}) to the form, 

\begin{eqnarray}
 \zeta(1-\zeta) S''(\zeta) + \left[2 l - \lambda - 1 + 2\zeta(a-l) - 2 a \zeta^2\right]\;S'(\zeta) + \left[\xi - a (\lambda + 1) - l(l-1) + 2 l a - 2 a l \zeta\right] S(\zeta) = 0, \label{se1_z}
\end{eqnarray}
where 
\begin{eqnarray}
l = \frac{\lambda + 2}{2} \pm \frac{1}{2}\sqrt{\lambda^2 - 4},  \qquad \xi = \frac{E}{2\hbar^2} + \frac{a^2}{4} - 1. 
\label{lvalue}
\end{eqnarray}

Equation (\ref{se1_z}) is in the form of the confluent Heun equation \cite{el2013confluent}. To find its solutions, we adopt the functional Bethe-Ansatz method, a quasi-exact technique proposed to solve differential equations of the form:
\begin{equation}
\sum_{j=0}^{4} a_j \zeta^j S''(\zeta) + \sum_{j=0}^{3} b_j \zeta^j S'(\zeta) + \sum_{j=0}^{2} c_j \zeta^j S(\zeta) = 0,
\label{bethe}
\end{equation}
where $a_0, a_1, a_2, a_3, a_4, b_0, b_1, b_2, b_3, c_0, c_1,$ and $c_2$ are parameters, \cite{zhang2012exact, agboola2012unified}. 

In this method, this equation (\ref{bethe}) is assumed to admit a polynomial solution of degree $n$ as, 
\begin{equation}
S(\zeta) = \prod_{i=1}^{n} (\zeta - \zeta_i), \quad \text{with } S(\zeta) = 1 \text{ for } n = 0,
\label{st}
\end{equation}
where $\zeta_1, \zeta_2, \dots, \zeta_n$ are distinct roots. These roots satisfy the Bethe-Ansatz equations:
\begin{equation}
\quad\sum_{j \neq i}^{n} \frac{2}{\zeta_i - \zeta_j} = -\frac{b_3 \zeta_i^3 + b_2 \zeta_i^2 + b_1 \zeta_i + b_0}{a_4 \zeta_i^4 + a_3 \zeta_i^3 + a_2 \zeta_i^2 + a_1 \zeta_i + a_0}.
\label{bethe-ansatz}
\end{equation}

These solutions hold under the following parameter constraints:
\begin{eqnarray}
c_2 &=& -n(n-1)a_4 - n b_3, \label{con1}\\
c_1 &=& -n b_2 - n(n-1)a_3 - (2(n-1)a_4 + b_3) \sum_{i=1}^{n} \zeta_i, \label{con2}\\
-c_0 &=& (2(n-1)a_4 + b_3) \sum_{i=1}^{n} \zeta_i^2 + 2 a_4 \sum_{i < k}^{n} \zeta_i \zeta_k + (2(n-1)a_3 + b_2) \sum_{i=1}^{n} \zeta_i + n(n-1)a_2 + n b_1.  \label{con3}
\end{eqnarray}

By comparing the equation (\ref{se1_z}) with (\ref{bethe}), we can obtain the eigenfunctions, 
\begin{equation}
\psi_{n}(r, \phi) = e^{i\lambda_n \phi} r^{\lambda_n} (1 + r^2)^n \; \exp\left[\sqrt{\frac{A}{2\hbar^2}}\frac{1}{1 + r^2}\right] \times \prod^{n}_{i= 1} \left(\frac{1}{1 + r^2} - \zeta_i\right), \quad 0 < r \leq \infty \label{solution}
\end{equation}
with energy eigenvalues
\begin{equation}
E_n = -\frac{A}{4} + 2\hbar^2 + \hbar \sqrt{2 A} \left[2 \sum^{n}_{i = 1}\; \zeta_i  + \lambda_n + 1\right], \label{energy}
\end{equation}
for the value,  $\lambda$, 
\begin{equation}
\lambda_n = -\frac{1}{n+1}\left(n^2 + 2 n + 2\right), \quad n = 0, 1, 2, ..., \label{lambda_value}
\end{equation}
and the roots $\zeta_i, i = 1, 2, 3, \dots, n$ can be evaluated from 
\begin{equation}
\sum_{j \neq i} \frac{2}{\zeta_i - \zeta_j} = \frac{2 a \zeta^2_i - 2 (n + a) \zeta_i + 2 n + \lambda_n + 1}{\zeta_i(1 - \zeta_i)}
\label{root}
\end{equation}

From the solutions (\ref{solution}) and (\ref{energy}), the ground state solution  is given by 

\begin{eqnarray}
\psi_{0}(r, \phi) &=& \frac{e^{-2i\phi}}{r^{2}}\;\exp\left[\sqrt{\frac{A}{2\hbar^2}}\frac{1}{1 + r^2}\right], \quad 0 < r \leq \infty,\nonumber\\
E_0 &=& -\frac{A}{4} + 2\hbar^2 - \hbar \sqrt{2 A}, \label{groundstate}
\end{eqnarray}
 where $\lambda_0 = -2$, (vide Eq. (\ref{lambda_value})). 
The first excited state,  where $\lambda_1 = -5/2$, can be expressed as 
\begin{eqnarray}
\psi_{1}(r, \phi) &=& \frac{e^{-5\;i\;\phi/2}}{r^{5/2}}\;(1 + r^2)\exp\left[\sqrt{\frac{A}{2\hbar^2}}\frac{1}{1 + r^2}\right]\left(\frac{1}{1 + r^2} - \zeta_1\right), \quad 0 < r \leq \infty, \nonumber\\
E_1 &=& -\frac{A}{4} + 2\hbar^2 + \hbar \sqrt{2 A} \left(2 \zeta_1 - \frac{3}{2}\right) \label{excitedstate1}
\end{eqnarray}
with ${\displaystyle \zeta_1 = \frac{1}{2} + \sqrt{\frac{\hbar^2}{2 A}} \pm\frac{1}{2} \sqrt{\frac{2\hbar^2}{A} + 1 + \sqrt{\frac{2\hbar^2}{A}}}}$. 

Equations (\ref{solution}) - (\ref{excitedstate1}) represent the quasi-exact solutions for the ground and first excited states of a one-dimensional spin chain Hamiltonian in the absence of a magnetic field, derived using the pre-canonical quantization method. The Holstein-Primakoff transformation maps spin degrees of freedom to bosonic modes using harmonic oscillator ladder operators,  capturing quantum fluctuations through the creation and annihilation of magnons \cite{holstein1940field}. These solutions (\ref{solution}) and (\ref{energy}) demonstrate that a conformal mapping of the Bloch sphere onto the complex plane suggests a nonlinear oscillator representation for quantum spin-wave excitations. Furthermore, the study generalizes these findings to incorporate both in-plane and out-of-plane motion, thereby establishing a connection between classical dynamics and quantum mechanics.

\subsection{Off-Plane motion:}
The out-of-plane $(S^x - S^z)$ motions are characterized by the Hamiltonian (\ref{H_PQ}) when $Q = 0$ and $B =0$, 
\begin{eqnarray}
\mathcal{H}(P) = \frac{1}{2}(1+P^2)^2 \Pi_P^2 - \frac{A}{4} \frac{(1-P^2)^2}{(1+P^2)^2}, 
\label{hamiltonianHP}
\end{eqnarray}
which is  integrable classically \cite{daniel1992singularity}. 

The quantum dynamics of this Hamiltonian can be analyzed using the pre-canonical equation:
\begin{eqnarray}
i \hbar \frac{\partial \Psi(P, z)}{\partial z} = \left[\frac{1}{2\;\sqrt{m(P)}} \hat{\Pi}_{P} \frac{1}{\sqrt{m(P)}}\hat{\Pi}_{P} - \frac{A}{4} \frac{(1 - P^2)^2}{(1 + P^2)^2} \right]\Psi(P, z),  \; 
\label{se_p}
\end{eqnarray}
where ${\displaystyle m(P) = \frac{1}{(1 + P^2)^2}}$ and $\hat{\Pi}_P$ is the conjugate momentum operator as expressed as  $\hat{\Pi}_P =-i\;\hbar\frac{\partial}{\partial P}$ and the Hamiltonian is ordered in non-hermitian ordered form \cite{carinena2004non}. As in the previous case. We solve the equation (\ref{se_p}) by considering 
\begin{equation}
\Psi(P, z) = \exp{\left(-\frac{i}{\hbar} E z\right)} \psi(P), 
\end{equation}
where $E$ is the energy eigenvalue, which reduces the equation (\ref{se_p}) to

\begin{eqnarray}
(1 + P^2)\frac{d}{dP}(1 + P^2)\frac{d\psi(P)}{dP} + \left(\frac{A}{2\hbar^2} \frac{(1 - P^2)^2}{(1 + P^2)^2} + \frac{2E}{\hbar^2}\right) \psi(P)= 0. 
\label{se_p1}
\end{eqnarray}

In general, equation (\ref{se_p1}) describes a curvature dependent nonpolynomial nonlinear oscillator, as discussed in \cite{V_2021}. This system can be analyzed using the Bethe-Ansatz method,  which is a quasi-exact treatment and imposes constraints on the quantization of the Hamiltonian (\ref{hamiltonianHP}). Hence, arbitrary choices of the ordered forms of the corresponding Hamiltonian operator are restricted.

In the following, we examine the complete solvability of the system (\ref{hamiltonianHP})  for the particular choice of ordering between  ${\hat \Pi}_P$ and $m(P)$, as specified in equation (\ref{se_p}). 

By applying the transformation, $ P = \cot{\frac{\theta}{2}}$, equation (\ref{se_p1}) can be simplified to 
\begin{eqnarray}
\frac{d^2 \psi}{d\theta^2} + \left(\frac{E}{2\hbar^2} + \frac{A}{16 \hbar^2} + \frac{A}{16\hbar^2} \cos(2\;\theta)\right)\psi = 0. 
\label{se_z}
\end{eqnarray}
Equation (\ref{se_z}) is in the form of the well-known Mathieu differential equation \cite{gradshteyn2014table, mclachlan1947theory, richards1983mathieu} which read as
\begin{equation}
w''(x) + \left(a - 2 q \cos(2 x)\right) w(x) = 0, \qquad ' =\frac{d}{dx}, \label{mathieu_eqn}
\end{equation}
which admits the solution
\begin{equation}
w(x) = C_1 Ce(a, q, x) + C_2 Se(a, q, x), \label{mathieu_soln}
\end{equation}
where $C_1$ and $C_2$ are arbitrary constants, $\mathrm{Ce}(a, q, x)$ and $\mathrm{Se}(a, q, x)$ are known as Mathieu functions. Mathieu functions can exhibit a range of behaviors depending on the parameters, $a$ and $q$. They may be periodic and bounded, non-periodic yet bounded, or even unbounded. For $q = 0$, these reduce to trigonometric functions, $\mathrm{Ce}(a, 0, x) = \cos(\sqrt{a}\,x), \quad \mathrm{Se}(a, 0, x) = \sin(\sqrt{a}\,x).$  When $q \ge 0$, the solutions can be periodic in $x$, with period $\pi$ or $2\pi$, and may be expressed as infinite Fourier series of cosine or sine terms. These periodic solutions are known as Matheiu functions of first kind and also known as the even and odd Mathieu functions $\mathrm{Ce}_{2n}(q, x), \mathrm{Ce}_{2n+1}(q, x), \mathrm{Se}_{2n}(q, x), \mathrm{Se}_{2n + 1}(q, x)$ with eigenvalues, $a_{2n}, a_{2n+1}, b_{2n}, b_{2n+1}$ for $n = 0, 1, 2, ...$. These functions are bounded and square-integrable on the interval $(0, 2\pi)$ \cite{gradshteyn2014table}. 

Further, Matheiu functions for non-zero value of $q$ and for the same value of $a$ can be expressed as, \cite{ mclachlan1947theory, richards1983mathieu}, 
\begin{equation}
Ce_{\nu}(q,x) = \cos(\nu x) + \sum^{\infty}_{i = 1} q^{i} C_i(x), \quad Se_{\nu}(q, x) = \sin(\nu x) + \sum^{\infty}_{i = 1} q^{i} S_i(x) \label{solution_nu}
\end{equation}
and correspond to discrete eigenvalues $a_{\nu}$,
\begin{equation}
a_{\nu} = \nu^2 + \sum^{\infty}_{i = 1} \alpha_i q^{i},  \label{characteristic-value}
\end{equation}
which depend on $q$. Here $\nu$ is a fractional number. The functions in equation~\eqref{solution_nu} are periodic when $\nu $ is a rational number, and non-periodic but bounded when $\nu$ is irrational. The coefficients $ C_i(x)$, and $S_i(x)$ can be expressed in terms of cosine and sine functions characterized by $\nu$, and also the parameters $\alpha_i$ are expressed in terms of $\nu$ \cite{mclachlan1947theory}.

On comparing equation (\ref{se_z}) with (\ref{mathieu_eqn}), and the parameters  
\begin{equation}
a = \frac{E}{2\hbar^2} + \frac{A}{16 \hbar^2}, \qquad q = -\frac{A}{32\hbar^2}, \label{parameters_value}
\end{equation}
we can now express the general periodic solution \cite{mclachlan1947theory, richards1983mathieu} as 
\begin{equation}
\psi_{\nu}(\theta) =  C_1 Ce_{\nu}\left(q, \theta\right) + C_2 Se_{\nu} \left(q,  \theta\right) \label{general_solution1}
\end{equation}
and the corresponding energy eigenvalues become
\begin{equation}
\hspace{1cm}\;E_{\nu} = -\frac{A}{8} + 2\hbar^2\left[\nu^2 + \sum^{\infty}_{i = 1} \alpha_i q^{i}\right].  
\end{equation}
The rational and irrational nature of the  $\nu$ values determines whether the bounded eigenfunctions (\ref{general_solution1}) are periodic or non-periodic. 

Hence, the quantum version of the one-dimensional spin chain system (\ref{hamiltonianHP}) becomes completely solvable, analogous to its classical counterpart.

\subsection{In-plane configuration}
The Hamiltonian (\ref{H_PQ}) has also been analyzed classically in the case \( P^2 + Q^2 = 1 \), which describes the in-plane motion of the system, \cite{daniel1992singularity}. In this case, the motion is restricted to the \( S^x - S^y \) plane and takes the following form:
\begin{eqnarray}
H(P, Q) &=& 2\left(\Pi_P^2 + \Pi_Q^2\right) + \frac{\mu B}{4} P,
\label{ham_bloch}
\end{eqnarray}
where the anisotropy term is excluded due to the motion being confined to the plane. The conjugate momenta are defined as:
\begin{equation}
\Pi_P = \frac{P_z}{4}, \qquad \Pi_Q = \frac{Q_z}{4}. 
\label{Pi_bloch}
\end{equation}
The classical analysis of the corresponding equation of motion leads to a solution that can be expressed using hyperbolic functions, resembling the static form of the pulsed soliton in the isotropic case.

We understand that the Hamiltonian (\ref{ham_bloch}) is simple and not complex, as the mass function is reduced to a constant. Now we use the pre-canonical quantization method to study the quantum behavior of the system. The corresponding equation is written as, 
\begin{equation}
\hat{\mathcal{H}}(P, Q)\Psi(P, Q, z) = i\hbar\frac{\partial\Psi(P, Q, z)}{\partial z}. \label{pre-canonicaleq}
\end{equation}

On quantization, we consider that 
${\displaystyle \hat{\Pi}_P = {-i}{\hbar} \frac{\partial}{\partial P}}$ and ${\displaystyle \hat{\Pi}_Q = {-i}{\hbar} \frac{\partial}{\partial Q}}$. 
Then we apply the transformation 
\begin{equation}
\Psi(P, Q, z) = \exp{\left(-\frac{i}{\hbar} E z\right)} \psi(P, Q
)
\end{equation}
and reduce the equation (\ref{pre-canonicaleq}) to
\begin{equation}
\left[-2\hbar^2\left( \frac{d^2}{dP^2} + \frac{d^2}{dQ^2} \right)+ \frac{\mu B}{4} P\right] \psi(P, Q) = E \psi(P, Q). \label{offplane_system}
\end{equation}

But when the equation (\ref{offplane_system}) is re-expressed in polar coordinates, $ P = \cos{2 \phi} $ and $ Q = \sin{2 \phi} $, it becomes the Mathieu differential equation (\ref{mathieu_eqn})
\begin{equation}
\left[\frac{d^2}{d\phi^2} - \frac{\mu B}{2\;\hbar^2} \cos{2\;\phi} + \frac{2\;E}{\hbar^2}\right] \psi(\phi) = 0.\label{se_phi}
\end{equation}

On comparing equation (\ref{se_phi}) with (\ref{mathieu_eqn}), and the parameters  
\begin{equation}
a = \frac{2E}{\hbar^2}, \qquad q = \frac{\mu B}{4\hbar^2}, \label{parameters_value}
\end{equation}
we can now express the periodic solution for (\ref{offplane_system}) as 
\begin{equation}
\psi_{\nu}(\phi) =  C_1 Ce_{\nu}\left(q, \phi\right) + C_2 Se_{\nu} \left(q,  \phi\right), \label{general_solution}
\end{equation}
where $C_1$ and $C_2$ are arbitrary constants and $\mathrm{Ce}_{\nu}(q, x)$ and $\mathrm{Se}_{\nu}(q, x)$ are the Mathieu functions (vide Eq. (\ref{solution_nu})). The energy eigenvalues are obtained from (\ref{characteristic-value}) as 
\begin{equation}
\hspace{1cm}\;E_{\nu} = \frac{\hbar^2}{2}\left[\nu^2 + \sum^{\infty}_{i = 1} \alpha_i q^{i}\right], \label{energy-inplane} 
\end{equation}
which depend on $q$. Here, $\alpha_i$'s are determined by imposing periodicity condition on the Mathieu functions. 

For $q = 0$, it means that $ B = 0 $ from equation (\ref{parameters_value}), and the characteristic value becomes $ a = n^2$, where $n$ is an integer. This fixes the energy eigenvalues to be 

\begin{equation}
\hspace{2cm}\;E_{n} = \frac{\hbar^2}{2} n^2, \qquad n = 0, 1, 2, 3, \dots.  \label{energy-inplane} 
\end{equation}
The eigenfunctions (\ref{general_solution}) then take the form, 
\begin{equation}
\hspace{-1cm}\psi_n(\phi) =  C_1 \cos(n \phi). \label{general_solution_q0}
\end{equation}
These solutions are obviously periodic \cite{richards1983mathieu}. 

Hence, we can express the solution for the pre-canonical equation \label{pre-canonical_phi} in the absence of the magnetic field as 
\begin{eqnarray}
\hspace{2cm}\Psi_{n}(P, Q, z) = C_1 e^{\frac{i}{\hbar} E_n z}\cos{\left[\frac{n}{2} \arctan\left(\frac{Q}{P}\right)\right]}, \quad n = 0, 1, 2, 3,...  \label{solution_final}
\end{eqnarray}
with energy eigenvalues given in Eq. (\ref{energy-inplane}). 

The Schr\"{o}dinger equation (\ref{se_phi}) can also be analyzed numerically. 

\section{Conclusion}
This work aims to extend the integrability aspects of the Heisenberg spin chain model into the quantum regime. Previous classical studies of the one-dimensional anisotropic Heisenberg spin chain in a transverse magnetic field, in the continuum limit, demonstrated that the system is integrable in the absence of either anisotropy or a magnetic field. Painlev\'{e}'s singularity structure analysis was applied to the modified Landau-Lifshitz equation, expressed in terms of a complex field variable resulting from the stereographic projection of the spin vector onto the complex plane \cite{daniel1992singularity}.

This analysis produced a Hamiltonian density for the static case and we identified it as a nonlinear sigma model. However, due to the complexity of quantizing the non-standard Hamiltonian density within infinite-dimensional configuration spaces, we used a local quantization scheme known as pre-canonical quantization method. This method treats all field variables and their conjugate momenta on equal footing within a ``multi-temporal framework" \cite{kanatchikov2001precanonical}. Consequently, it results in a generalized Schr\"{o}dinger equation for the field wave function.

Using the Bethe-Ansatz method, we determined that the system is quasi-exactly solvable. Furthermore, we analyzed the quantum solvability of the system for two specific cases: in-plane motion and off-plane motion. We derived the admissible solutions, that is the eigenfunctions and their corresponding energy eigenvalues.

This work thus provides valuable insights into the integrability and quantum solvability of this continuum model.

\section*{AUTHOR’S CONTRIBUTIONS}		
Both the authors contributed equally. 

\section*{Acknowledgment}
V. C. would like to thank SRM TRP Engineering College, India, for their financial support, vide number SRM/TRP/RI/005. The research work of M. L. is supported by a SERB National Science Chair and he wishes to express his thanks to the Anushandan National Research Foundation, Department of Science and Technology, Government of India for the award.

\section*{Data Availability Statement}

The data supporting the findings of this study are available from the authors on reasonable request.

\end{document}